\documentclass{article}

\usepackage{arxiv}

\usepackage[utf8]{inputenc} 
\usepackage[T1]{fontenc}    
\usepackage{hyperref}       
\usepackage{url}            
\usepackage{booktabs}       
\usepackage{amsfonts}       
\usepackage{nicefrac}       
\usepackage{microtype}      
\usepackage{cleveref}       
\usepackage{lipsum}         
\usepackage{graphicx}
\usepackage{natbib}
\usepackage{doi}
\usepackage{soul}
\usepackage[usenames,dvipsnames]{color}

\title{Digital assistant in a point of sales}

\newif\ifuniqueAffiliation
\uniqueAffiliationtrue

\ifuniqueAffiliation 

\author{ 
    {Emilia Lesiak} \\
	Orange Innovation Poland\\
	\texttt{emilia.lesiak2@orange.com} \\
	\And
    {Grzegorz Wolny} \\
	Orange Innovation Poland\\
	\texttt{grzegorz.wolny@orange.com} \\
	\And
    {Bartosz Przybył} \\
	Orange Innovation Poland\\
	\texttt{bartosz.przybyl@orange.com} \\
	\And
    {Michał Szczerbak} \\
	Orange Innovation Poland\\
	\texttt{michal2.szczerbak@orange.com} \\
}

\else

\fi

\renewcommand{\shorttitle}{\textit{arXiv} Template}

\hypersetup{
pdftitle={Digital assistant in a point of sales},
pdfsubject={cs.HC},
pdfauthor={Emilia Lesiak, Grzegorz Wolny, Bartosz Przybył, Michał Szczerbak},
pdfkeywords={Digital Assistant, Voice User Interface (VUI), Customer Engagement, Multilingual Support, Experimental Analysis, Technological Adaptability, Customer Service Innovation, Retail Point of Sales},
}

\begin{document}
\maketitle

\begin{abstract}
This article investigates the deployment of a Voice User Interface (VUI)-powered digital assistant in a retail setting and assesses its impact on customer engagement and service efficiency. The study explores how digital assistants can enhance user interactions through advanced conversational capabilities with multilingual support. By integrating a digital assistant into a high-traffic retail environment, we evaluate its effectiveness in improving the quality of customer service and operational efficiency. Data collected during the experiment demonstrate varied impacts on customer interaction, revealing insights into the future optimizations of digital assistant technologies in customer-facing roles. This study contributes to the understanding of digital transformation strategies within the customer relations domain emphasizing the need for service flexibility and user-centric design in modern retail stores.
\end{abstract}


\section{Introduction}

The rapid advancements in artificial intelligence (AI) are reshaping various industries, with telco operators being significant beneficiaries. This sector, crucial for global connectivity, faces the dual challenge of escalating customer expectations and the need to remain competitive in a swiftly evolving market. To address these challenges, there is a growing reliance on innovative technologies such as chatbots, voicebots, and videobots. The integration of these digital tools is viewed as a strategic response to enhance customer interactions, streamline operations, and maintain market relevance.

Among these technologies, digital assistants equipped with Voice User Interfaces (VUIs), including those with graphical screen displays, are gaining prominence. These interfaces promise to make customer service interactions more natural and engaging. This study focuses on the deployment of a fully animated digital character, acting as an assistant powered by a voice interface, at a sales point. The aim is to assess its impact on user satisfaction, engagement, and problem-resolution efficacy.

Our study was guided by the following research questions:

1: How do digital assistants influence customer engagement and problem-resolution outcomes?

2: What are the challenges and opportunities associated with the use of VUI technologies in practical service environments?

By exploring these questions through a systematic experimental setup, this paper seeks to contribute new insights into digital transformation strategies within customer service. It aims to evaluate the effectiveness of digital assistants and explore their potential in enhancing the quality of service delivery. This introduction sets the stage for detailed discussions on methodology, findings, and implications for future innovations and best practices in customer relations channels. 

The article explores the application of a digital assistant in a retail setting, examining the roles and integration of artificial intelligence and VUIs within digital assistants. It starts by setting the technological context, followed by a detailed description of the experimental design, including the deployment, data collection methodologies, and analysis procedures. The results section discusses the impact and implications of these technologies in service environments, highlighting both challenges and opportunities discovered during the experiment. Finally, the paper concludes by summarizing key findings and offering recommendations for further research and practical application with a focus on improving customer service through digital assistants.

\section{Related Work}

Voice-User Interfaces, which first emerged in the 1990s, represent a significant advancement in human-machine interaction, facilitating direct voice communication between users and systems. The history of VUIs reflects their evolution from simple task automation, like call routing, to complex interaction management through sophisticated natural language processing, thereby enhancing user experience and operational efficiency. Notable early systems include AT\&T's voice recognition call processing which adeptly directed calls based on voice commands\cite{Wilpon1990}. The development of advanced systems such as Spoken Dialogue Systems (SDS) and Embodied Conversational Agents (ECAs) marked a significant expansion in the scope of VUIs. These systems integrate speech with other modalities such as body language and facial expressions to create more engaging and natural interactions. Studies like those by \cite{cassell2000s} and \cite{breazeal2004designing} detail how these integrations enhance the immersiveness of human-computer interactions, a crucial aspect for user acceptance.

Artificial Intelligence (AI), Machine Learning (ML), and Natural Language Processing (NLP) have further propelled the capabilities of digital assistants, expanding their functionality to encompass a broad range of tasks including internet searches, schedule management, and smart device control. Digital assistants such as Siri, Google Assistant, Alexa, and Cortana exemplify this technological progression, providing invaluable assistance in daily activities and redefining user interaction with digital platforms \cite{luger2016like}, \cite{mctear2016conversational}, \cite{amershi2019guidelines}.

Research into VUI design emphasizes the necessity for systems to accurately understand and process natural language inputs while providing intuitive, contextually relevant responses. Challenges related to speech recognition accuracy, accommodating diverse user accents, and the naturalness of system-generated responses persist, highlighting the complexity of these systems. \citep{morgan1999build}, \cite{mctear2016conversational} and \cite{lopez2018alexa} provide a comparative analysis of speech recognition technologies, illustrating the technological advancements and remaining hurdles in achieving seamless interaction.

In customer service, digital assistants have demonstrated the potential to revolutionize service delivery by ensuring availability around the clock, reducing response times, and personalizing user interactions. Insights from \cite{xu2019waveear} show significant impacts on customer satisfaction and operational efficiency. Moreover, the studies by \cite{xie2020global} underline the importance of transparency, reliability, and security in building user trust and acceptance, pivotal for positive customer experiences.

Emerging trends in digital assistants development, such as the integration of multimodal interfaces and the application of decentralized technologies like blockchain, point towards a future where digital assistants are not only more capable and secure but also better tailored to diverse user needs \cite{oviatt2018handbook}, \cite{mik2019smart}. \cite{velkovska2019emotional} highlights the potential of emotionally aware characters that recognize and respond to human emotions, enhancing the quality of customer relationships. The latter work describes also the closest experiment to the one described in this document.

Studies could further benefit from a deeper exploration of ethical considerations in development of digital assistants, especially as these systems become more autonomous and integrated into everyday life. \cite{van2018methodological} discuss the importance of ethical transparency, particularly around nudging techniques in consumer interactions. A more extensive discussion on these ethical issues would provide a comprehensive view of the responsibilities of developers and designers in this field.

In conclusion, this review underscores the vital integration of advanced technology and human-centric design in the development of voice-user interfaces for digital assistants. As the society progresses deeper into the age of artificial intelligence, it is crucial that VUIs not only enhance functionality but also prioritize ethical design, privacy, and emotional intelligence. The challenges identified here highlight the dynamic nature of the field and the opportunities for innovative research. Future efforts should focus on refining natural language processing and enhancing user engagement to develop digital assistants that are not just tools but trusted companions. This continuous advancement will help ensure that technology meets human needs with respect and integrity, shaping a more interconnected and empathetic future.

\section{Experiment Setup}

The integration of a digital assistant within a retail environment represents an advancement in merging technologies with voice user interfaces to enhance customer service and increase sales efficiency. This study was conducted in a high-traffic retail store known for its tech-savvy clientele and commitment to sustainability. This setup aimed to investigate the utility of the digital assistant in traditional retail settings, focusing on optimizing customer interactions across various service scenarios, including product inquiries and supporting sales. One might notice, that this setup, as well as the three months period adopted for the experiment have certain similarities to the setting in \cite{velkovska2019emotional}, where the researchers observed a robotic arm with a face visualization using also a VUI in a telco shop in Paris.

For our assistant, however, we used a three-dimensional digital character created using Unity 3D software. The character's body was styled as a futuristic robot with mechanical elements and smooth, white surfaces, emphasizing its modern and technological nature. We chose this digital character for our study because it has become the central figure in Orange Poland's advertising campaigns, significantly enhancing the brand's visibility and its connection with customers across various platforms. 

The assistant was fully animated, meaning its movements were fluid and realistic, and synchronized with speech, making communication with it more natural and engaging. Additional graphics and elements supporting interactions could appear on the screen, such as changing skins and artifacts, which could be adjusted depending on the research context. Utilizing motion sensors, the character could respond to the presence of clients, actively initiating interactions, such as waving in greeting. All these features made the digital character not only effectively support the conduct of research but also serve as an attractive and interactive element designed to increase user engagement.

\begin{figure}[htbp]
    \centering
    \begin{minipage}[b]{0.3\textwidth}
        \centering
        \includegraphics[width=\textwidth]{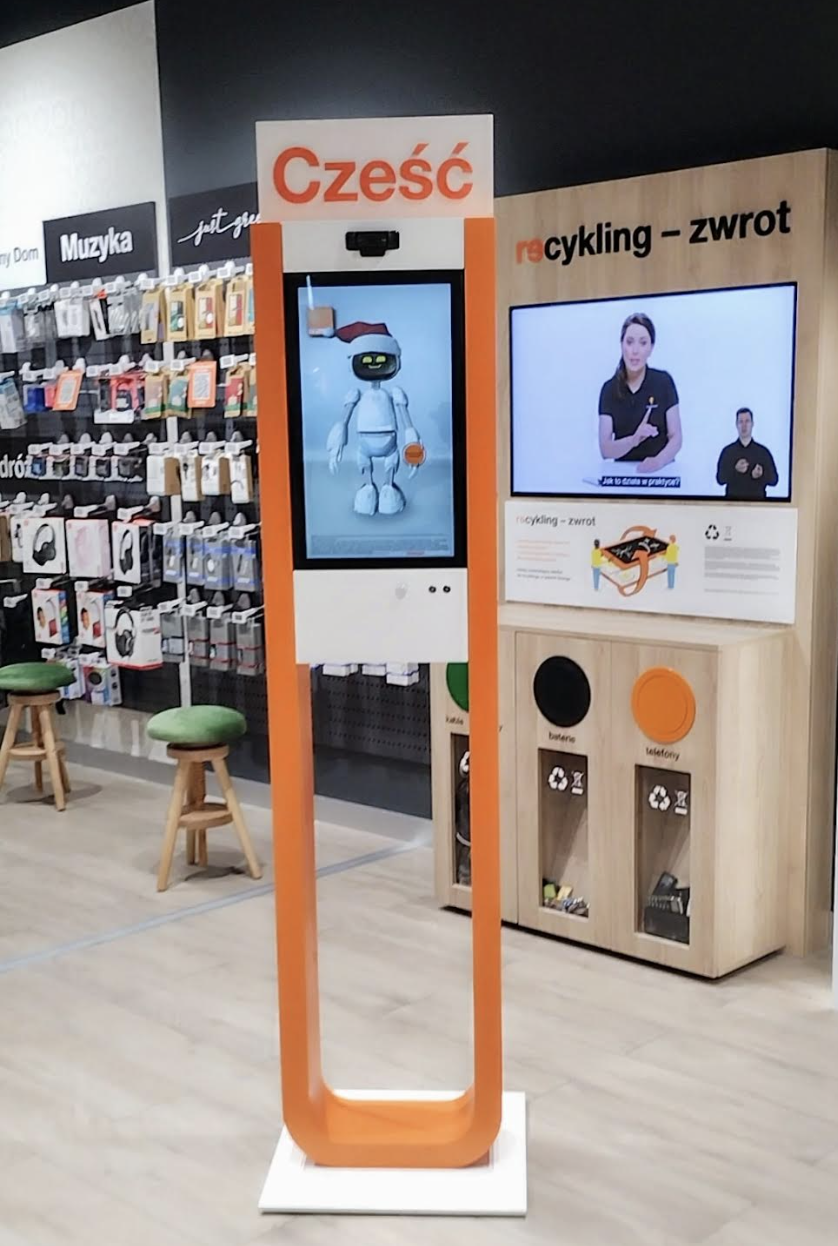}
        \caption{The stand in the point of sales}
        \label{fig:max-stand}
    \end{minipage}
    \hfill
    \begin{minipage}[b]{0.3\textwidth}
        \centering
        \includegraphics[width=\textwidth]{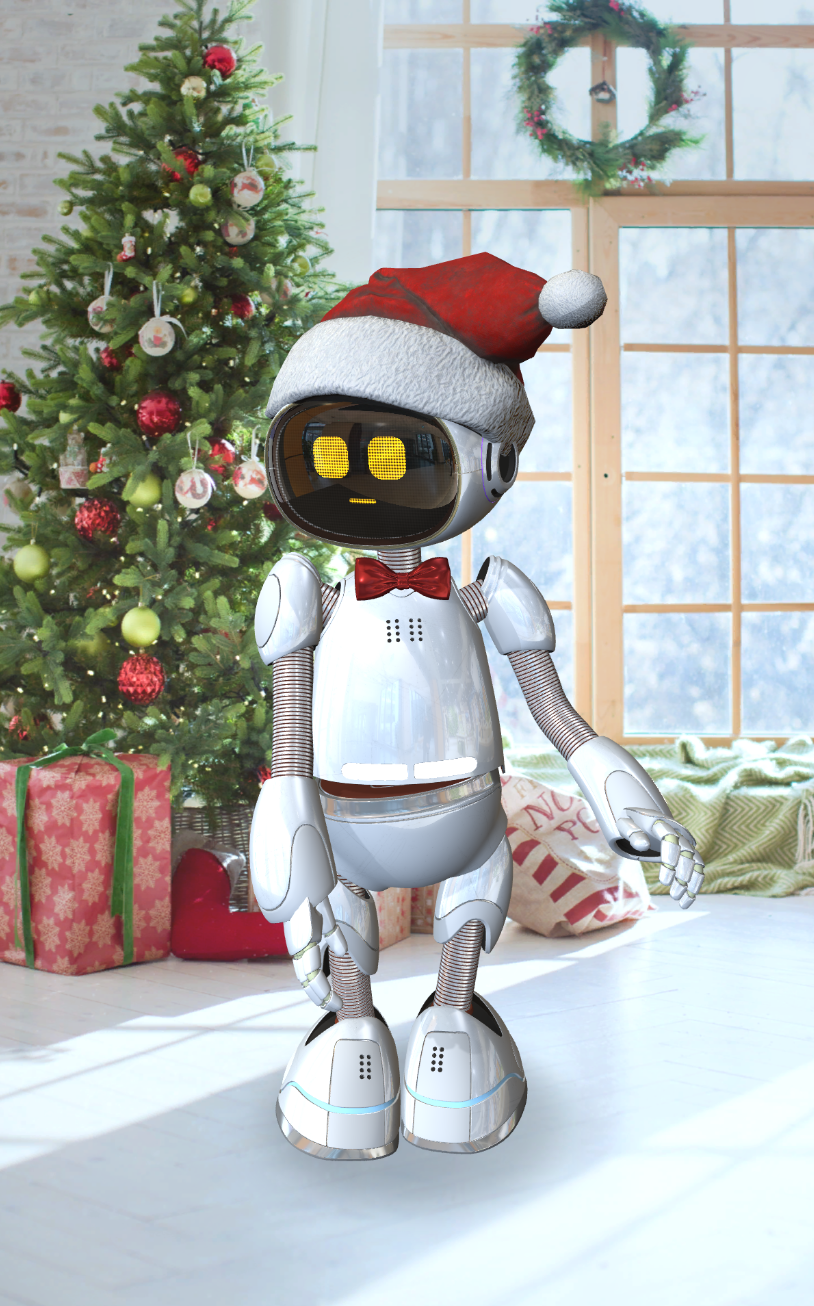}
        \caption{Christmas / winter "skin"}
        \label{fig:max-xmax}
    \end{minipage}
    \hfill
    \begin{minipage}[b]{0.3\textwidth}
        \centering
        \includegraphics[width=\textwidth]{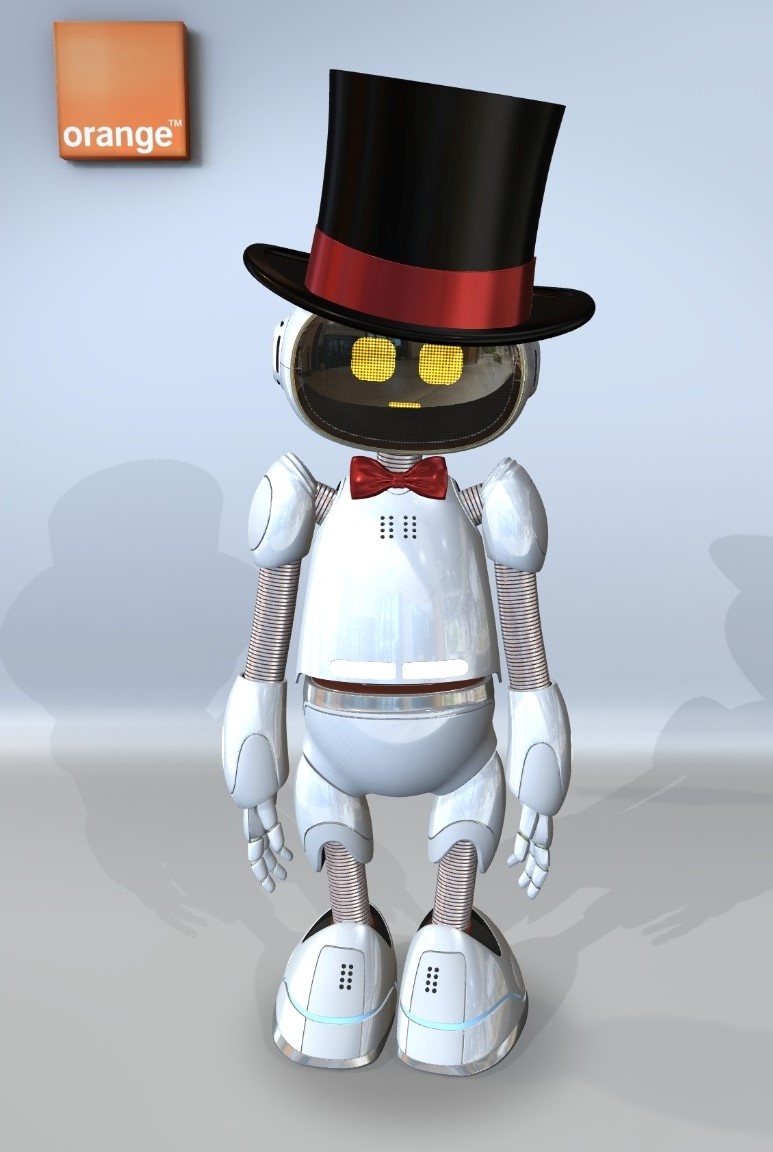}
        \caption{Alternative New Year's Eve "skin"}
        \label{fig:max-hat}
    \end{minipage}
\end{figure}

The digital assistant was tailored to align with the store’s operational strategies and customer service needs. The implementation process began with an in-depth analysis of business and customer needs, defining key functionalities such as conversation management, product presentation, and support during visits to the point of sales. Component selection was key to effectively support these features, ensuring the assistant was user-friendly and easily accessible.
The main functionalities introduced a set of features aimed at improving customer interactions:

\begin{itemize}
    \item Sales Pokes: Employing both verbal and visual cues, the assistant proactively informs customers about current and seasonal offerings. This feature was designed to engage customers in a two-stage process, initially capturing their attention with prompts and subsequently encouraging further exploration through detailed follow-ups.
\end{itemize}
\begin{itemize}
    \item Conversation Engine: At the core of the assistant is a powerful conversational interface, capable of handling a wide array of discussions, from specific sales details to broader customer service inquiries.
\end{itemize}
\begin{itemize}
    \item Phone Recommendations: A feature that enhances the shopping experience by offering personalized mobile phone recommendations based on individual user preferences and specific requests.
\end{itemize}
\begin{itemize}
    \item Multilingual Capabilities: The assistant's ability to communicate in multiple languages, including English, French, and Ukrainian, in addition to Polish, to enhance accessibility to a diverse demographic.
\end{itemize}
\begin{itemize}
    \item Feedback Loop: This functionality collects and analyzes user feedback, playing a crucial role in the ongoing development cycle of the assistant, facilitating continual enhancements based on user interactions.
\end{itemize}
The rollout timeline for these features spanned from December 1, 2023, to February 2024, structured to progressively introduce and enhance functionalities:

\begin{itemize}
    \item December 2023: Implementation of multilingual support and the "poke - learn more" feature to improve user engagement and accessibility.
\end{itemize}
\begin{itemize}
    \item January 2024: Introduction of the personalized phone recommendation feature and graphical updates to the assistant’s avatar, enhancing the visual and interactive appeal.
\end{itemize}
\begin{itemize}
    \item February 2024: Deployment of targeted sales pokes and refined conversation prompts aimed at boosting interaction quality, coupled with incentives for deeper exploration of the assistant’s capabilities.
\end{itemize}
The systematic implementation of these features aimed to gradually improve the user experience by increasing usability, accessibility and engagement. The gradual introduction of new elements also contributed to increasing analytical and reasoning capabilities, providing knowledge on the possibilities of improving and optimizing the functionality of the assistant.

\section{Assistant's design}

This section describes the technical architecture of the voice user interface-based digital assistant and its hardware and software components that were prepared in the context of our study. We introduce those by explaining first the user journey during the interactions with the assistant.

\subsection{User journey}

A person who enters the premises of the point of sales is detected by the assistant, which waves or greets the customer depending on the circumstances. An intrigued user who wants to talk to the assistant approaches the stand. He or she is given the opportunity to change the language from the default Polish using the graphical interface through touching a corresponding national flag pictogram for French, English or Ukrainian.

Then one needs to press and hold a push-to-talk button to ask a question, releasing the button when finished. At that point, the question is analyzed and the assistant expresses that process with its eyes taking the form of turning gears. After a moment the user can hear the assistant's response and see the corresponding visual effects, like the model's animation, and eventually an image with an offer that the user may click on to get more information.

The help offered by the assistant is mostly of an informational nature, it can explain different telco-related terms, talk about the offer, and eventually cover some general knowledge topics or weather. Should a user have any requests requiring actions or user identification, like performing actual sales or dealing with after-sales service, it redirects to the human staff. In cases of improper handling the push-to-talk button, the assistant gives necessary instructions.

The conversation can be ended by simply walking away, which the assistant is capable of detecting, but also through the process of a conversation. After a customer declines any further assistance or says "goodbye", an invitation to evaluate the experience is proposed by the assistant. The evaluation on the 1 to 5 scale can be performed either vocally or using the touchscreen showing images of 5 empty stars.

\subsection{Architecture}

The above functionalities required conceiving and implementing both the front-end physical stand, an interface to the customers, and the back-end set of software services. Both communicate with each other using a network connection. 

\paragraph{Input/output interface}
The assistant is exposed to users using an almost 2-meter-high box with a 55 centimeters diameter touch screen in the upper part, the top edge of the screen situated at 179 centimeters from the ground as shown in Figure \ref{fig:max-stand}. The casing hides the hardware parts, including the central processing unit and cables, whereas the screen displays the assistant's avatar in a form of a three-dimensional animated robot. Its model moves and gestures according to its activity, namely: waiting for interactions, processing user requests in the back-end, talking or expressing different built-in "emotions".

The stand is equipped with a generic webcam-class microphone and speakers to allow for voice interactions. The video from the camera is not captured. The speakers are a generic off-the-shelf product. In our case the interactions were also tactile for some clickable images with offer, but also for the push-to-talk functionality. Indeed, we have decided to use a "press and hold" interface for the P2T, as alternatives like "press once to talk" or "no button" solutions would require more complex audio signal processing, i.e. speech and silence detection, which we evaluated as too risky in a noisy environment.

Furthermore, the stand used two types of sensors, namely infra-red and ultrasonic, helping in detecting user presence in a wide spectrum in front of the installation and in measuring the distance between the two interacting parties respectively. This gave the assistant its "eyes" in the absence of a camera, which we were restricted from employing due to legal reasons.

\paragraph{On-site software}
The robot model, its animations and the scene was powered by an application created in Unity framework. The sensors were coupled with a logic for firing some predefined robot actions, like waving when someone is supposedly passing by or greeting on someone standing still. Other software components handled also the data related to the interactions, bridging thus the user interface with the back-end services.

Moreover, the physical stand and its software services could be accessed remotely, allowing us to easily and quickly monitor important technical parameters, troubleshoot issues, and most importantly, manage updates to the experimental functionality, including the introduction of new features according to the study timeline.

\paragraph{Interactions data flow}
The back-end processed the data using a number of steps. First the audio data was converted to text, which was then passed to the conversational engine and processed there. The engine, which managed the dialogue between the assistant and the user, was returning textual responses. Those texts were converted to audio and sent back to the stand. The audio-text conversions were mostly dealt with using existing cloud services, only the test-to-speech for Polish language required an additional open-source software adapted for the experiment.

The conversational engine we employed was the Rasa X Community Edition\footnote{\href{https://rasa.com/rasa-x/}{https://rasa.com/rasa-x/}} based on the Rasa Open Source component. We had provided the content for the engine by defining possible user queries, known as intents, and responses for them, which resulted in an AI model to handle conversations. The textual content for possible questions was provided in Polish only since the underlying BERT model for extracting features is able to deal with any language due to its multilingual nature. 

We have also extended the conversational engine with the ability to query external services for some of the intents, namely Wikipedia for general knowledge inquiries and a weather forecast provider. Finally, The dialogue management engine could also return additional data, which was interpreted by the application on the stand, e.g. an image to display or a particular animation to be run.

Whereas the model used in this environment managed messages in any language, we needed to supply the translations to other languages for any required pre-defined replies. Those that were generated on the fly, for external data sources in particular, required an additional on-demand translation service.

\paragraph{Data storage}

For the purpose of the study, the system was designed to store conversation texts, data from the sensors mounted on the stand, system diagnostic data from the stand and other events. All data was stored in an InfluxDB instance as it allows to store numerical and text data in the time series format and provides rich data query capabilities. Finally, for efficiency and service responsiveness reasons, a text-to-speech cache was introduced to avoid audio generation each time for repeated assistant lines.

Due to the fact of conversations data being potentially quite sensitive, we have put much attention on assuring the privacy of the interactions. First of all, as already mentioned, we have decided not to employ any cameras for "seeing" the user - instead, we have restricted the solution to the motion and distance sensors. Secondly, the voice of the customer, which can be her or his digital signature, is deleted directly after it is transformed into text. And lastly, any remaining personal data in the conversation text, should a user provided any names or identifiers, is also automatically anonymized.

\section{Experiment results}
\label{sec:experiment_results}

In this section, we present the results based on the data obtained during the 3 months of the experiment. Indeed, our approach to evaluate the experiment differs from the one adopted by \cite{velkovska2019emotional}, where the researchers employed a human observer of the interactions between shop clients and their robot assistant as well as feedback surveys. We, on the other hand, opted for relying on the analyses of the real usage data, while the in-person study was rather a part of the design process preceding the actual experiment.

We begin by discussing the data sources utilized (\ref{sec:exp_datasources}). Subsequently, we delve into various aspects grouped into three areas. Subsection related to initial user engagement (\ref{sec:exp_user_engagement}) presents metrics concerning the user journey to learn about the digital assistant's presence ranging from foot traffic at the point of sales, through the number of assistant reactions triggered by the movement sensor, time spent in front of the assistant, the number of initiated interactions, until we reach the analysis of interactions in different languages. The next subsection addresses observations related to the actual interaction experiences (\ref{sec:exp_interactions}) by giving insights into data related to conversation duration, the flow of the conversation and technical issues. The last part focuses more on the business perspective (\ref{sec:exp_business}) and presents results from the analysis of topics addressed and the impact of attention-grabbing pokes on sales levels.

\subsection{Data sources}
\label{sec:exp_datasources}
In our study, we harnessed a diverse array of data sources, including low level data stored in databases. These event-based records were time-series in nature, capturing the temporal dimension of interactions between the virtual assistant and customers. The data encompassed a variety of interactions, including dialogue exchanges, button presses, screen touches, notification displays, and survey responses.

Moreover, our analysis incorporated data from a distance sensor equipped on the assistant. This sensor collected distance measurements at intervals of 0.5 seconds, providing a continuous stream of data reflecting the proximity of customers to the virtual assistant. These distance measurements were logged enriching our dataset with real-time insights into customer proximity and engagement with the virtual assistant.

Additionally, we utilized logs from the conversational system built upon the Rasa framework to gain deeper insights into the dialogue dynamics between the virtual assistant and customers. These logs offered a granular view of conversational flows, user intents, and system responses, enabling us to evaluate the effectiveness of the virtual assistant in addressing customer inquiries and providing assistance.

Furthermore, we integrated external data on foot traffic within the store premises, as well as across other locations. This facilitated an analysis of customer visitation patterns over time, allowing us to discern any shifts in foot traffic dynamics following the introduction of the virtual assistant.

In conjunction with these datasets, we examined sales performance metrics disaggregated across different product categories. By correlating sales data with the aforementioned data sources, we sought to ascertain the impact of the virtual assistant on purchasing behavior and its contribution to variations in sales volumes across various product lines.

By leveraging the above-mentioned diverse datasets our analysis aimed to provide a comprehensive understanding of the efficacy of the virtual assistant in enhancing customer experiences and driving sales within the retail environment. Note that some of the presented charts, i.e. those related to foot traffic and sales levels have intentionally been stripped of specific numerical values due to the sensitivity of the data being presented.

\subsection{User engagement}
\label{sec:exp_user_engagement}

Our analysis delves into several facets of the background and initial stages of user engagement with the virtual assistant.

\begin{figure}[htbp]
    \centering
    \begin{minipage}[b]{0.49\textwidth}
        \centering
        \includegraphics[width=\textwidth]{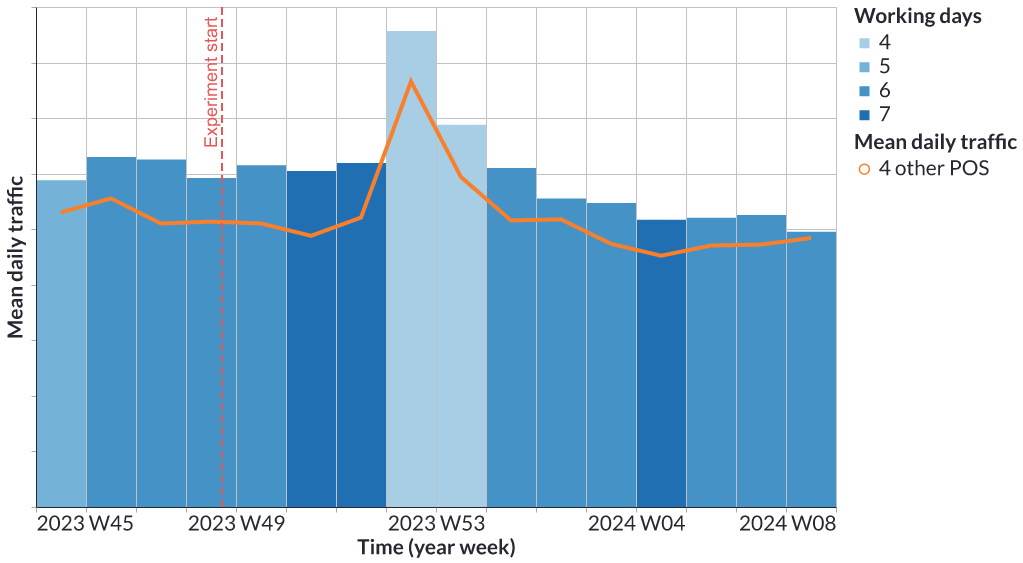}
        \caption{Traffic level in point of sale}
        \label{fig:traffic}
    \end{minipage}
    \hfill
    \begin{minipage}[b]{0.49\textwidth}
        \centering
        \includegraphics[width=\textwidth]{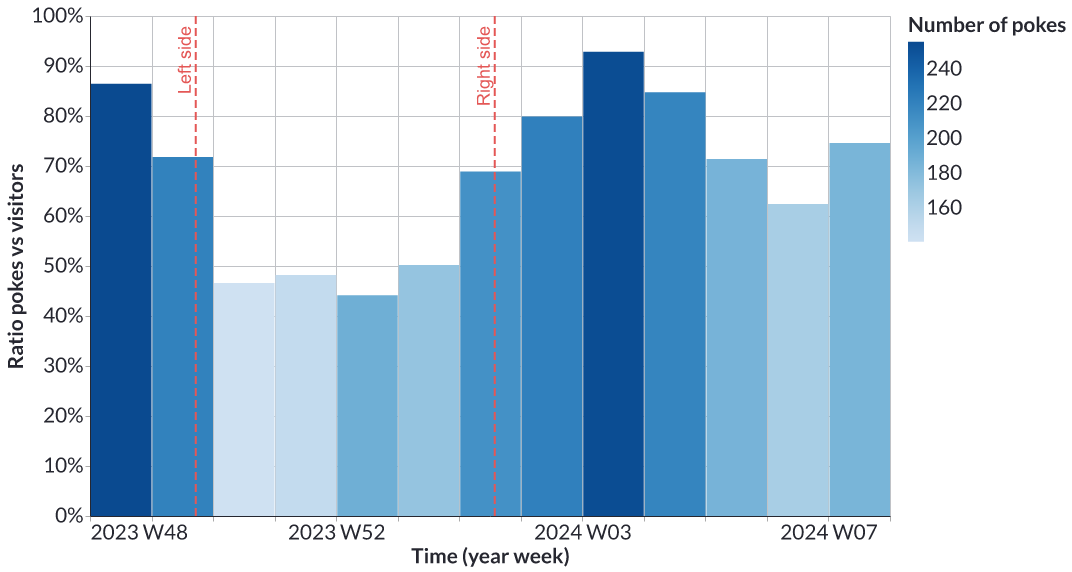}
        \caption{Level of pokes}
        \label{fig:pokes}
    \end{minipage}
    \vspace{1cm}
    \vfill
    \begin{minipage}[b]{0.49\textwidth}
        \centering
        \includegraphics[width=\textwidth]{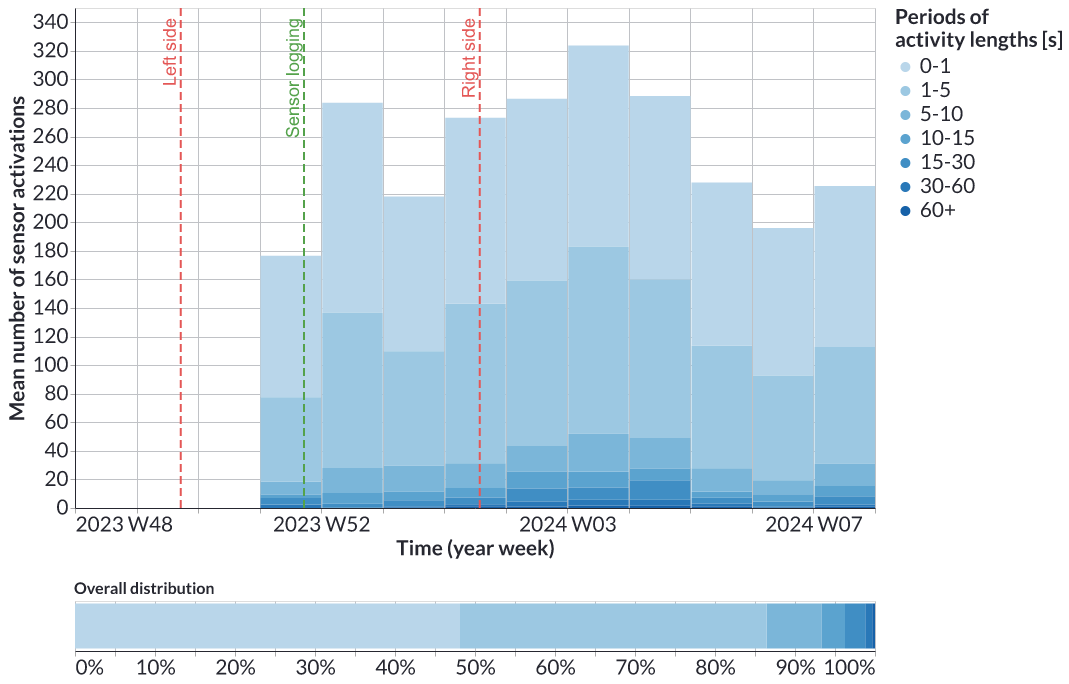}
        \caption{Activity time from sensors}
        \label{fig:sensors}
    \end{minipage}
    \hfill
    \begin{minipage}[b]{0.49\textwidth}
        \centering
        \includegraphics[width=\textwidth]{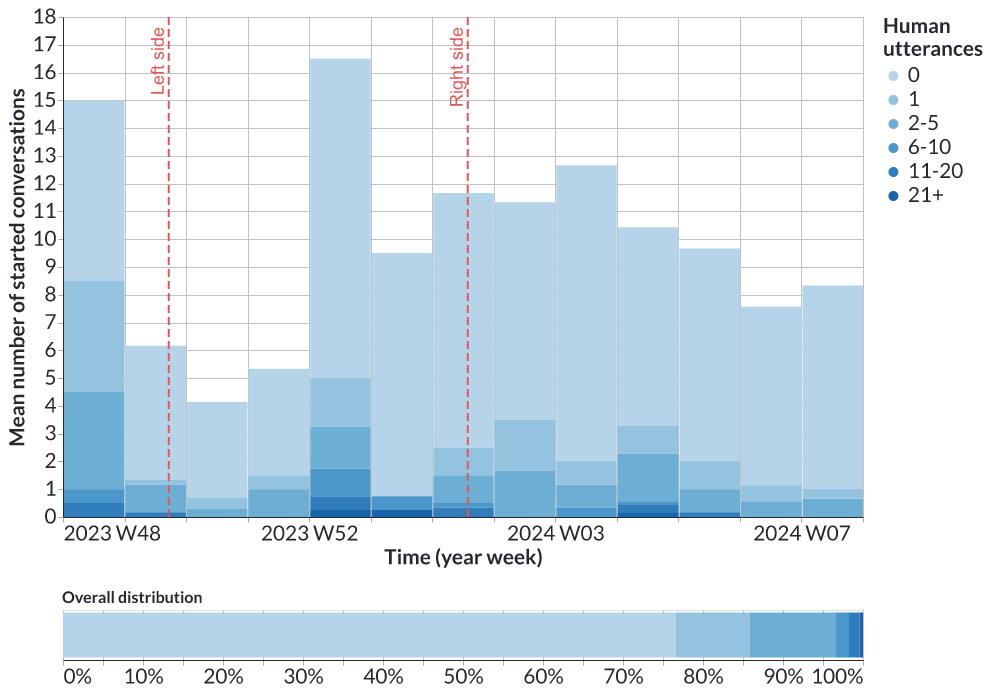}
        \caption{Level of interactions}
        \label{fig:interactions}
    \end{minipage}
    \vspace{1cm}
    \vfill
    \begin{minipage}[b]{0.49\textwidth}
        \centering
        \includegraphics[width=\textwidth]{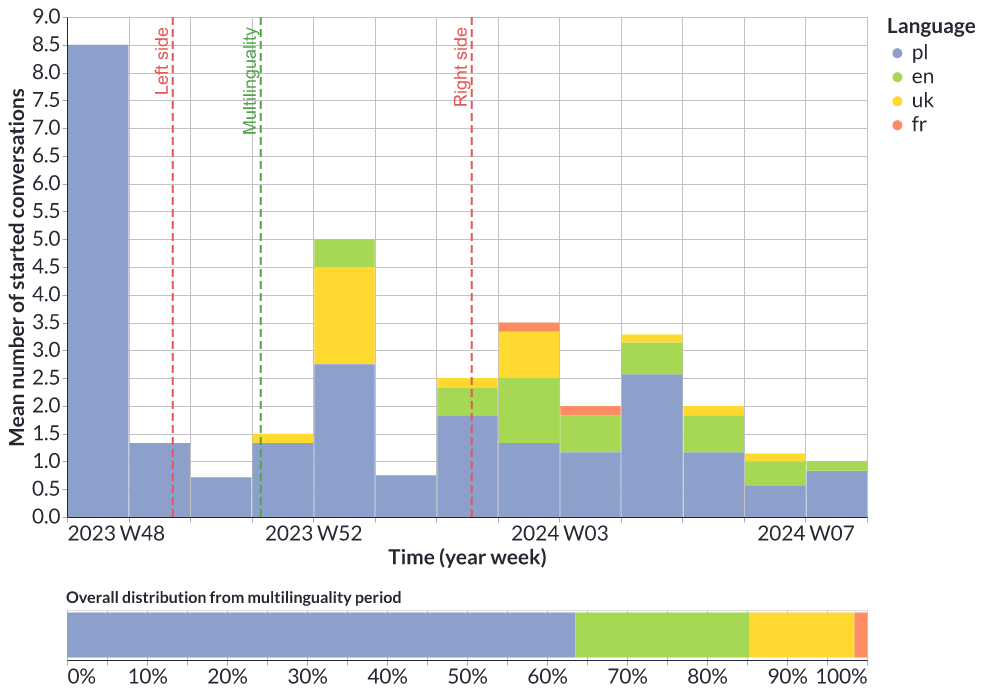}
        \caption{Conversations by languages}
        \label{fig:languages}
    \end{minipage}
    \hfill
    \begin{minipage}[b]{0.49\textwidth}
        \centering
        \includegraphics[width=\textwidth]{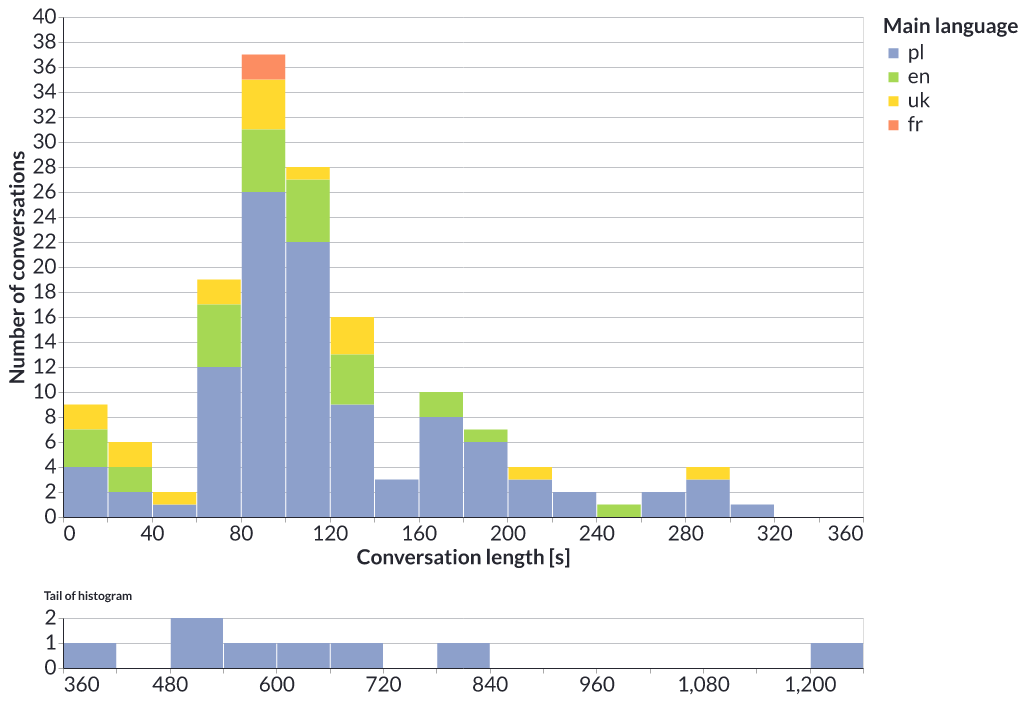}
        \caption{Conversations lengths}
        \label{fig:hist_length}
    \end{minipage}
\end{figure}

\paragraph{Traffic level}

The analysis of daily visitor traffic aims in bringing a perspective into the impact of our experiment into the point of sales routines. Figure~\ref{fig:traffic}, illustrates the average daily footfall in the store from November 2023 onwards, segmented on a weekly basis. It is evident that the level of foot traffic has remained relatively stable before and after the commencement of the VUI based digital assistant experiment in December 2023 and that its introduction did not significantly alter the store's visitation patterns.

Comparing the foot traffic trends of the analyzed store with the average of four other similar stores, represented by the orange line, reveals marginally higher values in the subject store. However, the overall behavior of the analyzed metric, characterized by fluctuations and trends, remains highly analogous across all stores. Even the peak towards the end of the year, driven by the holiday season and fewer working days in the two peak weeks, occurred in both the analyzed case and the others. This suggests that factors influencing the dynamics of the traffic level are likely consistent and irrespective of the presence of the digital assistant.

\paragraph{Level of pokes}

Figure~\ref{fig:pokes} presents the number of vocally and visually attention-grabbing elements, referred to as \textit{pokes} activated upon user motion detection. An important factor here is that the virtual assistant changed its location twice during the observation period. Initially positioned on the right side of the store, close to the mall corridor, it was then relocated to the left side, farther away from the main aisle, before returning to the right side but positioned deeper within the store.

The graph illustrates a notable variation in the percentage of displayed pokes relative to the number of visitors, depending on the virtual assistant's placement. In the locations on the right-hand side, this percentage fluctuated between 60\% to over 90\%, with an average of around 77\%. Conversely, when situated on the left, the percentage did not exceed 50\%, with an average of approximately 48\%.

\paragraph{Time spent in front of assistant}

In order to better examine the whole journey of the client entering a shop, passing before the digital assistant and eventually engaging in an interaction or not, we focus now on the motion sensor data that started to be collected after approximately three weeks into the experiment. These data were gathered at half-second intervals, providing an approximation of the distance from the sensor and enabling an examination of the duration of time individuals spent in proximity to the virtual assistant, and depicted in Figure~\ref{fig:sensors}.

Our analysis considered intervals shorter than 2 seconds, which stand for 4 measurements, within a continuous sequence of measurements to be part of the same interaction. Results indicate that nearly half of the activations were momentary, that is below 1 second, indicating instances where individuals were passing by the virtual assistant without pausing. Approximately 40\% of activations fell within the range of 1 to 5 seconds, suggesting momentary halts in front of the virtual assistant and potential engagement with it. Longer pauses exceeding 5 seconds occurred at a rate of approximately 25-30 per day when the virtual assistant was positioned on the left side of the store, increasing to over 40 per day in peak period when it was placed on the right side.

\paragraph{Started interactions}

Another step in the engagement process is the physical interaction with the robot assistant, starting from the moment a conversation is initiated by pressing the "Start conversation" button or selecting a language. The average daily count of interactions for most analyzed weeks ranged between 7 and 12 but could drop to 4 during the weakest performing week and rise to 16 during the week between Christmas and New Year's Day - see Figure~\ref{fig:interactions}.

Shades of blue color denote the number of utterances from the human side. As evident from the graph and the distribution of interactions with different utterance counts, approximately 76\% of cases recorded no utterances from the person. Interactions with at least one utterance, referred to as \textit{dialogues}/\textit{conversations} henceforth, accounted for about 24\% of cases, thus averaging 2 to 4 conversations per day. These observations further corroborate the findings from the previous section.

\paragraph{Multilinguality}

The multilingual support was introduced in the fourth week of the experiment and was another planned mean of encouraging users to engage in conversations in other languages than Polish. This option was presented in the graphical interface using flag icons. The chart in Figure~\ref{fig:languages} illustrates the average daily number of conversations categorized by language. Additional languages included English, Ukrainian, and French. Analysis of the data reveals that approximately 36\% of conversations took place in a foreign language following the introduction of multilingual support, particularly in English and Ukrainian.

\subsection{Interaction experience}
\label{sec:exp_interactions}

This subsection focuses on analyzing user interactions with the assistant in the context of various factors. The first paragraph presents the analysis of conversation lengths. It then discusses the frequency of interruptions of assistant utterances by users, providing insights into the dynamics of the dialogue. The subsequent paragraph analyzes the effectiveness of intent detection in user utterances, identifying areas for improvement. Furthermore, technical issues encountered during the experiment are discussed, as well as the assessment of user satisfaction with the assistant interaction.

\begin{figure}[htbp]
    \begin{minipage}[b]{0.49\textwidth}
        \centering
        \includegraphics[width=\textwidth]{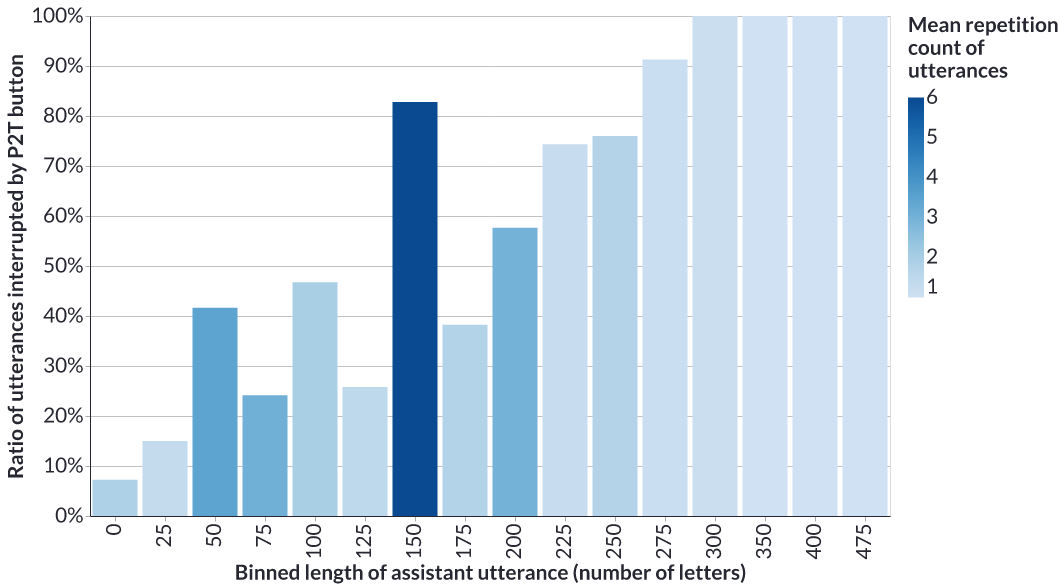}
        \caption{Interruptions of assistant utterances}
        \label{fig:p2t}
    \end{minipage}
    \hfill
    \begin{minipage}[b]{0.49\textwidth}
        \centering
        \includegraphics[width=\textwidth]{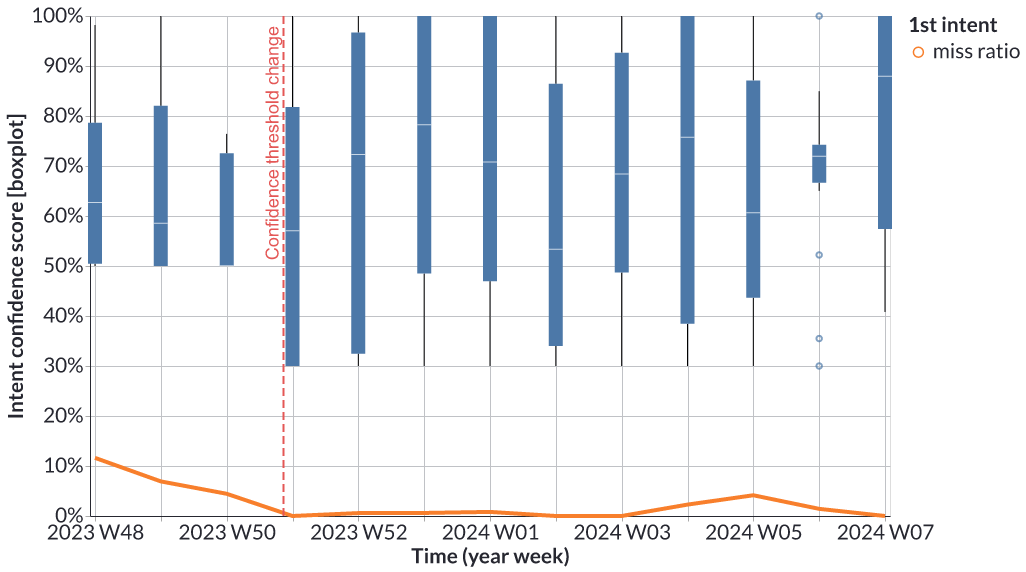}
        \caption{Intent detection confidence}
        \label{fig:confidence}
    \end{minipage}
    \vspace{1cm}
    \vfill
    \begin{minipage}[b]{0.49\textwidth}
        \centering
        \includegraphics[width=\textwidth]{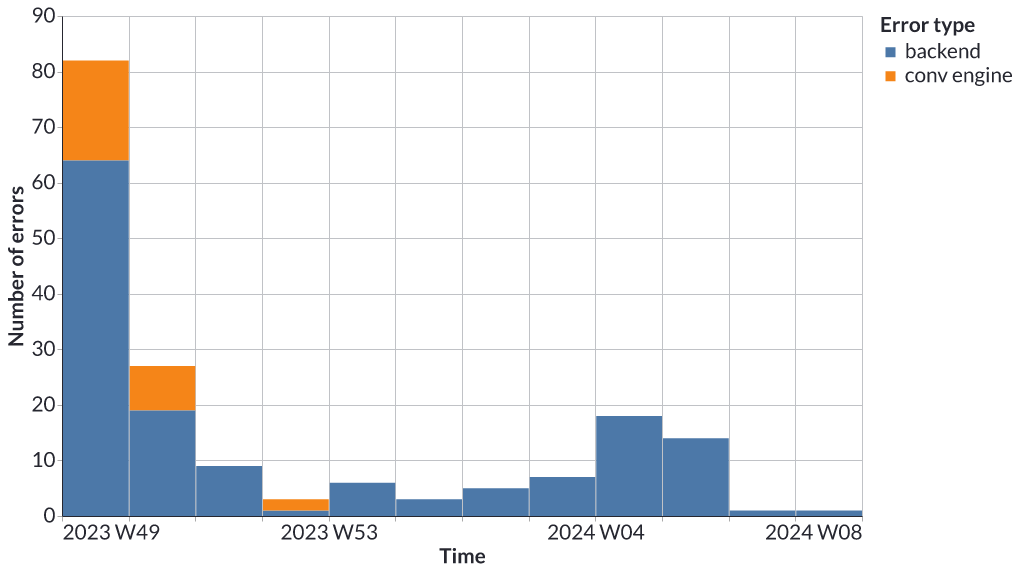}
        \caption{Number of errors in time}
        \label{fig:errors}
    \end{minipage}
    \hfill
    \begin{minipage}[b]{0.49\textwidth}
        \centering
        \includegraphics[width=\textwidth]{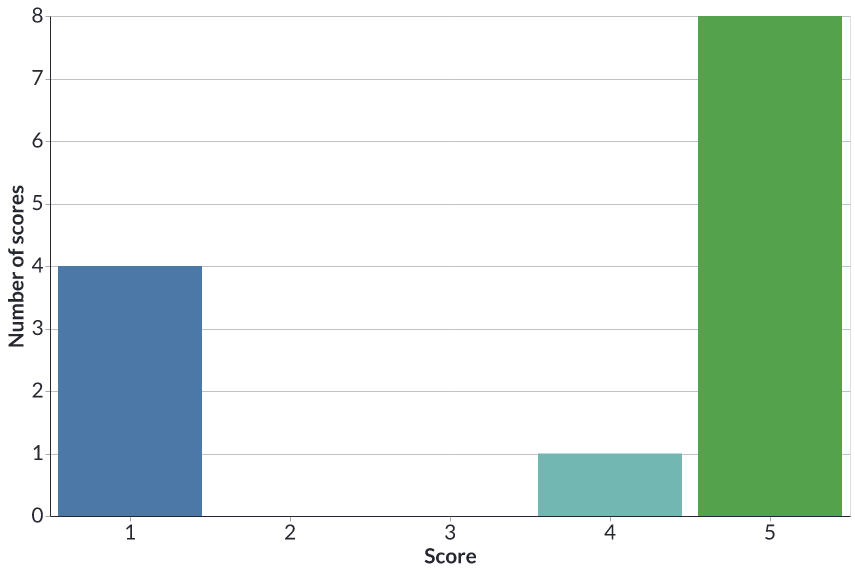}
        \caption{Survey scores}
        \label{fig:scores}
    \end{minipage}
\end{figure}

\paragraph{Conversations lengths}

One straightforward way of quantifying the experience is related to the duration of conversations. Figure~\ref{fig:hist_length} presents a histogram depicting this metric, with different colors standing for the four languages introduced. Analysis of the chart indicates that the majority of conversations lasted between 60 and 140 seconds. Only a few conversations, all of which were in Polish, extended beyond 5 minutes. Within the timeframe of less than 5 minutes, there were no significant differences in conversation lengths among different languages.

\paragraph{Interruptions of assistant utterances}

In order to better understand the way the people conversate with the digital assistant and the flow of the conversations, we took into consideration the timelines and the moments when people start to speak. As an interesting insight from this study, we present a chart in Figure~\ref{fig:p2t} which shows how often the assistant's utterances are interrupted (by pressing the P2T button) depending on their length measured by the number of characters. From the chart, one can see that utterances longer than 225 characters are interrupted in as much as 75\% of cases, with those longer than 300 characters never being fully listened to. 

The intensity of the blue color is correlated with the average number of utterance repetitions of the same length within the same conversation and provides an additional dimension to the chart. It shows that in the case of utterances repeated multiple times, the likelihood of interruption also increases. This is most evident, for example, in the case of an utterance serving as an instruction for using the P2T button, which is 155 characters long and corresponds to the dark blue bar on the chart. Its presence clearly indicates that a message repeatedly uttered during a single conversation was almost always interrupted by the user.

\paragraph{Human utterances}

We have also analyzed other available data points from logs generated by the assistant and subjectively evaluated the contents of the conversations themselves. Each conversation was assessed based on several criteria, including whether the utterances related to Orange-related topics, whether intents were correctly assigned, whether the speech-to-text module correctly recognized the text and whether the conversation could be considered satisfactory from both the user and the assistant perspectives.

Over the course of three months, 762 conversations were registered, of which 172 (22.6\%) contained at least one non-empty user utterance, totaling 637 utterances. Topics related to offers and the store were discussed in 87 (50.6\%) conversations and in 150 (23.5\%) utterances. For the Polish language, the percentage of conversations related to Orange was 44.4\%, while for English and Ukrainian, it was over 60\% (65.7\% and 61.1\%, respectively), highlighting the importance of support for non-native speakers, which should be developed with particular attention.

Figure~\ref{fig:confidence} depicts the confidence levels during the detection of intents in human utterances. Following three weeks of experimentation, the threshold level responsible for accepting the intent categorization was adjusted from 0.5 to 0.3. This adjustment resulted in a decrease in the miss ratio of the best-intent choice, as evaluated manually. Overall, less than 8\% of utterances were annotated as misses by the conversational engine. Furthermore, in nearly half of the cases where a miss occurred, the second choice was deemed a better option. These errors often arose in situations where multiple intents related to similar topics were present in the intent database. In such instances, a more effective approach would involve querying for additional details — a functionality that warrants consideration in future iterations.

\paragraph{Technical problems}

Throughout the duration of the experiment, we monitored errors occurring both on the conversational engine side and in the operation of the assistant itself, which might have also hindered user experience. The number of detected issues one can consult in the chart from Figure~\ref{fig:errors}. Even though the first week, the number of errors was quite high, it was quickly reduced significantly to below 10 errors per week for most of the time, and in the case of the conversational engine, even eliminated entirely. Additionally, we manually annotated suspicious cases of text recognition by the speech-to-text module. Less than 4\% of utterances were marked as incorrectly recognized, which is a very good result considering the assistant's environment, which can generate a lot of additional noise.

\paragraph{User satisfaction}

Without access to video and without analyzing audio sound for detected emotions, assessing user satisfaction during interactions with the assistant was a challenging task. One of the evaluation mechanisms was a feedback survey displayed when the assistant detected the end of the conversation or when the user explicitly expressed a desire to provide feedback. The collected ratings, ranging from 1 to 5, are presented in the chart in Figure~\ref{fig:scores}. Unfortunately, there are very few of them, only 13, with an average score of 3.6, but one can already observe a known tendency in declarative ratings hinting to more engagement in either very positive or very negative experiences \cite{park2018bias}. The survey was also displayed 6 additional times with no score given.

Additionally, we conducted manual annotation of conversations, assigning ratings of 0, 0.5, and 1 based on subjective satisfaction from the user's perspective and from the perspective of the assistant's performance and adherence to project assumptions. The conversation flow was evaluated, considering whether the user received substantive answers to their questions, whether they expressed signs of satisfaction or dissatisfaction, or whether they repeated similar questions several times, which could indicate that they did not receive the answer they were searching for. At the same time, in the latter case, as long as the repeated responses were provided in accordance with the predetermined key in the conversational engine, they were positively evaluated from the assistant's perspective. Similarly, this was the case in all instances where the assistant responded according to the designed operational schema. Summarizing this subjective annotation, we estimate that approximately 65\% of conversations can be considered satisfactory from a human perspective, while for the assistant ratings, this coefficient was 87\%.

\subsection{Business perspective}
\label{sec:exp_business}

Digital assistants are increasingly being integrated into various business environments to enhance customer interactions and streamline operations. In this section, we delve into the business implications of deploying a virtual assistant in a retail setting, focusing on key insights derived from conversation topics and sales data analysis.

\begin{figure}[htbp]
    \begin{minipage}[b]{0.49\textwidth}
        \centering
        \includegraphics[width=\textwidth]{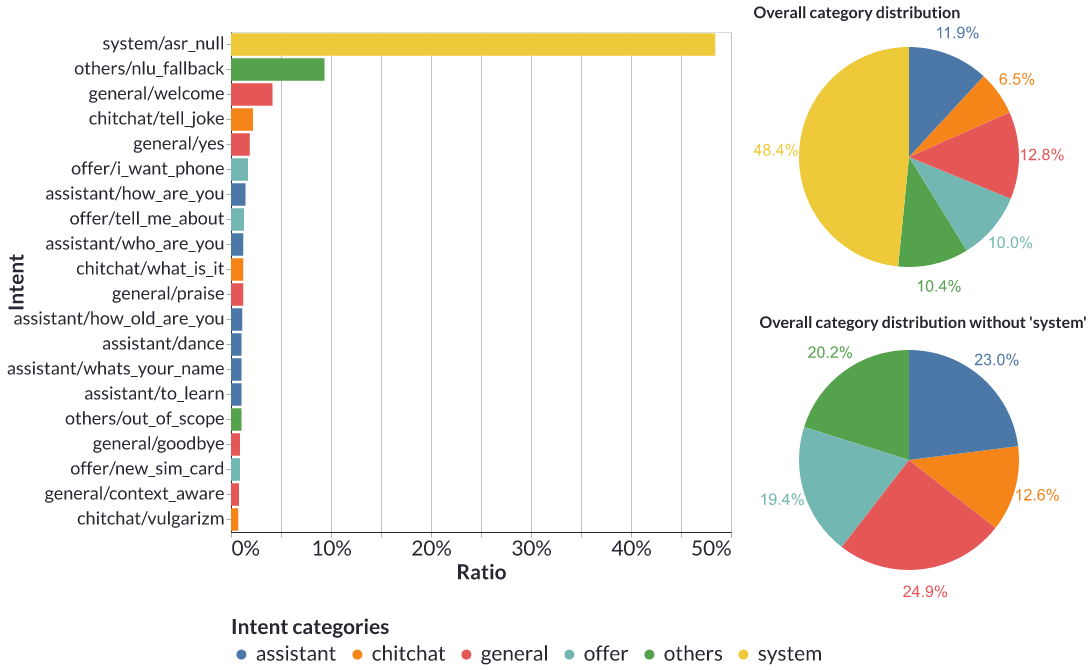}
        \caption{Most popular intents}
        \label{fig:intents}
    \end{minipage}
    \hfill
    \begin{minipage}[b]{0.49\textwidth}
        \centering
        \includegraphics[width=\textwidth]{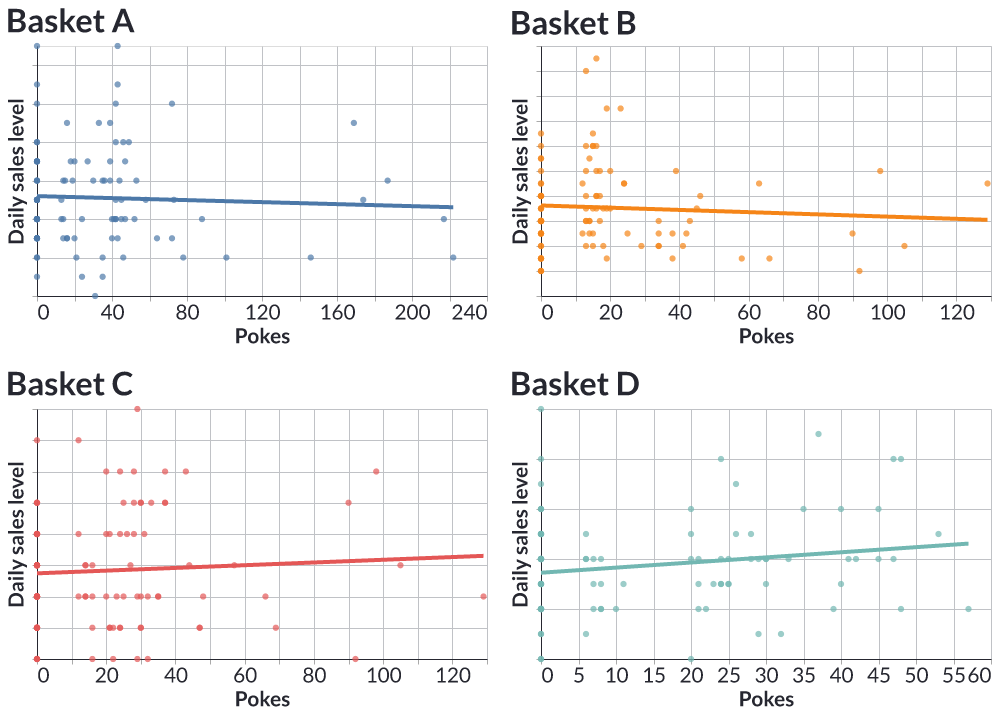}
        \caption{Impact of pokes on sales level}
        \label{fig:sales}
    \end{minipage}
\end{figure}

\paragraph{Most popular intents}

Based on meaningful two-sided conversations, we analyze the main intents of user utterances from our conversational engine. Each user utterance was assigned an intent from a previously prepared set. For the purpose of analysis, intents were grouped into the following categories:
\begin{itemize}
    \item \textit{offer} - utterances related to the company's offers, 
    \item \textit{general} - general utterances such as "\textit{yes}", "\textit{no}", "\textit{good morning}", 
    \item \textit{chitchat} - utterances related to queries such as weather, Wikipedia knowledge, time, jokes,
    \item \textit{assistant} - utterances related to the assistant itself, 
    \item \textit{others} - utterances for which it was difficult to provide an answer without access to external knowledge or due to difficulties in assigning intents.
\end{itemize}
A separate category of \textit{system} events, to which the assistant responded, was also identified. These events were generated by the assistant in response to certain user behaviors, triggering actions on the conversational engine side. Charts in Figure~\ref{fig:intents} show statistics related to the detected intents.

The analysis of these charts reveals that nearly half of the analyzed utterances were empty. This is a significant issue stemming from the communication solution used with the robot, which required users to press and hold the button while speaking to the assistant. This solution proved to be unintuitive and difficult for shop visitors to use, even despite the assistant's messages informing about the button usage upon detecting improper usage. 

A notable portion of utterances - category "others" could have been handled if the assistant could leverage generative AI and the power of LLMs. The decision not to utilize LLMs was dictated by the design choice of complete accountability to clients for the assistant's utterances in a point of sales setting, which, if generated by LLMs, could result in hallucinations, which were quite common at the time of the experiment. The remaining categories were roughly evenly distributed. It is worth noting the high percentage of utterances related to the assistant itself - users were interested in acquiring knowledge about it and its capabilities.

\paragraph{Impact of pokes on sales level}

One of the assistant's functionalities involved displaying promotional \textit{pokes}. Each of these was assigned to a specific product group offered in the store. Based on information about the number of pokes and sales levels in each basket, we analyzed whether the presentation of graphical incentives accompanied by voice messages had an impact on sales. The data is visualized in Figure~\ref{fig:sales}. Each point on the graph represents a single day and expresses the level of sales on the Y-axis and the number of displayed incentives on the X-axis, divided into four baskets for which we gathered the most information. Based on the analysis of these graphs, it is difficult to conclude that the information presented by the assistant influenced sales levels. 

An additional observation regarding the impact of pokes on users is the fact that the implemented "find out more" button functionality, an invitation to follow up on the promotion that accompanied each \textit{poke}, was not very popular, and it was used 49 times during the experiment.

\section{Results discussion}
\label{sec:results_discussion}

The experiment and the analysis of the collected data yield numerous insights, which we group into three complementary perspectives, namely: technical, user and business.

\subsection{Technical improvements}

Regarding the experiment described in \cite{velkovska2019emotional} the authors point out technical difficulties as one of the main pain points preventing the users to actually benefit from the assistant's help. Even though the technology, including AI models enhancing the audio processing in the VUIs, has significantly progressed since then, the technical context of the VUIs and conversational engines has still some margin of progress ahead.

Undoubtedly, the push-to-talk button proved to be a significant hurdle for many users. Despite graphical information (text on the button) and voice instructions provided by the assistant, the frequency of issuing these instructions indicates that such a mode of interaction was non-intuitive and challenging to master. It is difficult to imagine continuous background sound analysis as constant eavesdropping raises significant legal concerns and high costs for such a solution or triggering audio capture based on sound intensity making it difficult to filter in a noisy store environment. However, pressing the button at the beginning of a statement with automatic recognition of its conclusion and possibly enriching the graphical interface with an animation explaining how to use the voice interface could help mitigate or at least reduce the issue with P2T.

Regarding the development of the conversational engine, the analyses suggest that integration with generative AI language models could aid both in handling topics beyond the virtual assistant's domain and in diversifying the conversation. Data indicates that repetitive topics within the same conversation were often interrupted by the speakers. Analyses also revealed that people tended to interrupt long statements during the experiment, so this aspect should also be considered when designing the assistant's manner of speaking. Perhaps it is worth considering it as a factor that could be subject to personalization depending on the interlocutor's expectations. Ensuring conversational context is crucial. People naturally assume they can refer back to previous statements, which is not always easy to handle with intent-based conversational solutions. Here, Large Language Models capable of handling long contexts could come to the rescue. It is also important to adapt the technology used in the speech-to-text module so that it can contextually correct errors from the assistant business domain in translated statements.

The operation of the conversational engine based on the detection of each statement's intent revealed some problems. Firstly, the system had difficulties in categorizing intentions that were thematically close to each other. There were cases where two intents received a high confidence score, but neither exceeded the required threshold, and the assistant reacted as if it did not understand. To improve this aspect in the future, attention should be paid to intents that may be confused with each other or introduce functionality to inquire about details in case of uncertainty regarding categorization. The latter solution could increase the naturalness of the assistant's conversation. 

Additionally, it's not uncommon for user utterances to contain multiple intents, indicating a complexity that the assistant must navigate through effectively. The assistant was not prepared for this and reacted in its statement only to one of them, leaving the other unhandled or forcing repetition from the speaker. Improving keywords and named entities recognition is crucial, as evidenced by instances where the speech recognition module was finding the closest Polish words to company names of English origin. Enhanced accuracy in understanding user inputs is essential for a seamless conversational experience. Moreover, topics related to the context in which the digital assistant is working that are not yet explicitly addressed should be quickly identified based on conversations analysis, maybe even automatically, and where possible, handling of relevant threads should be continuously added. Such a proactive approach can contribute to enhancing user satisfaction and efficiency.

People frequently resort to using abbreviated forms rather than full phrases, opting for expressions like "\textit{Orange offer}", "\textit{Registration}", or "\textit{My Orange application}" instead of "\textit{could you tell me about ...}". This suggests a tendency towards brevity and efficiency in communication. Furthermore, users often operate under the assumption that the conversation context is maintained throughout the interaction. They expect the assistant to remember previous exchanges and tailor responses accordingly. Technical inquiries also feature prominently in user utterances, with individuals seeking details such as microphone positioning or the purpose of the P2T button. This underscores the importance of providing clear explanations and transparent functionality. Volume control emerges as a recurring topic of interest, reflecting users' desire for customization and control over their interaction experience. 

Interestingly, our observations suggest that younger users tend to exhibit more freedom in their interactions, exploring different functionalities and posing more general questions. Lastly, the entertainment function is a popular feature, with users frequently engaging in activities like enjoying jokes or requesting dances. For such users, the entertainment-enjoyment dimension had a significant role. Depending on the specific context of the assistant's operation, it is worth considering this aspect of its functionality. This highlights the importance of incorporating engaging and enjoyable elements into the assistant's repertoire as long as it fits the context and the role of the assistant in a given physical space.

\subsection{Value for the user}

The virtual assistant was designed to offer several valuable benefits to users, in order to enhance their overall experience and interaction satisfaction. One notable advantage is its multilingual capability, enabling users to conduct conversations and resolve issues in their preferred language. Analysis of user interactions revealed that individuals who spoke languages other than Polish more often raised inquiries related to Orange services, highlighting the importance of multilingual support in increasing service accessibility and inclusivity.

Furthermore, the virtual assistant extends beyond business domain matters to address broader topics and provide entertainment as a factor complementary to the straight-to-the-business approach. Users frequently engage in casual conversations, requesting jokes or even asking the assistant to dance. This additional functionality serves as a source of amusement, particularly for younger users, and helps alleviate waiting times in queues, enhancing the overall customer experience. This entertainment factor was reported, however, also a distractor in previous experiments (\cite{velkovska2019emotional}) and should be thus dosed with caution.

Moreover, the potential for personalization represents a significant value proposition for users. As the technology evolves, users may have the opportunity to customize the assistant's speech patterns or even enable automatic adaptation to their continuously detected needs. This level of personalization fosters a sense of tailored service and enhances user satisfaction, very critical to feel special rather than just as another customer a company puts an insensitive bot in front of. This is one of the aspects to experiment with in the future.

Additionally, the virtual assistant has potential to facilitate user feedback, allowing clients to express their opinions on the interaction experience but also on the point of sales service as a whole. This feedback loop not only enables users to voice their concerns but also can also provide valuable insights for service improvement efforts, ultimately leading to enhanced service quality and user engagement.

Another key benefit is the graphical presentation of conversation aspects, enabling users to better understand and visualize certain details discussed during the interaction, such as the option to view and compare available colors of a specific model of phone casing. This visual representation empowers users to make more informed decisions, enhancing their overall comprehension and decision-making process.

\subsection{Business perspective}

The deployment of virtual assistants in retail environments holds significant implications for business operations and customer service strategies. Despite their potential, several factors influence their effectiveness and integration into existing business frameworks.

One notable observation is the limited impact of virtual assistants on in-store foot traffic. Several possible reasons contribute to this phenomenon, including the assistant blending into the surrounding environment, lack of promotional activities to raise awareness of its presence, and pre-planned visits to the store. Understanding these dynamics is crucial for optimizing the deployment of virtual assistants and maximizing their influence on customer engagement.

The placement of virtual assistants within the retail environment plays a pivotal role in their effectiveness. In this experiment, some individuals mistook the assistant for a ticket machine, highlighting the importance of clear signage and intuitive user interfaces to attract attention and convey the assistant's function effectively. Strategic placement should be considered as part of the overall customer experience design, ensuring alignment with the assistant's intended role and functionalities.

For virtual assistants to effectively engage customers and prolong interaction duration, they must be capable of handling end-to-end use cases seamlessly. Frequent redirections to in-store personnel due to informational limitations hindered the depth of interactions and led to predominantly short and one-sided exchanges.

Multilingual support emerges as a critical aspect for reaching a wider audience. A significant proportion of conversations, particularly among non-Polish speakers, revolved around operational topics related to the store's offerings. By catering to diverse linguistic needs, virtual assistants can enhance accessibility and ensure effective communication with a broader customer base.

Moreover, the scalability of virtual assistants across multiple locations offers a consistent and reliable customer service experience. While initial implementation costs may be high, the potential for long-term cost savings in more repetitive in-store information, or sales routines as well as through value generation when extending work hours of human staff, underscores the value proposition of virtual assistants in retail settings.

Integration with existing systems and data collection mechanisms is paramount for maximizing the utility of virtual assistants. Long-term data collection provides valuable insights into customer preferences and behaviors, enabling data-driven decision-making to enhance sales effectiveness and customer satisfaction.

\section{Conclusion}

The study on the implementation of a digital assistant in a retail environment demonstrated its potential to improve customer service quality and operational efficiency, especially through multilingual support, which attracted user interest. However, the experiment revealed significant difficulties in capturing customer attention in retail spaces filled with digital advertising screens, which limited the assistant's effectiveness in maintaining consumer interest.

While the interactive functionalities enabled quick distribution of marketing information, their ultimate impact on sales was not significant. This suggests the need for more diverse promotional strategies. Technical challenges and a complicated user interface, especially the push-to-talk mechanism, hindered usability. These issues highlight areas requiring future improvements. Personalization and the integration of advanced technologies, such as large language models (LLMs), could enhance the assistant's usability and adaptability, increasing user satisfaction. 

To fully leverage the advantages of digital assistants in retail, continuous adjustments based on real-time data analysis and user feedback will be crucial. In the future, improving the technical capabilities of assistants, designing the user interface that focuses on the clients' needs as well as their end-to-end journeys through the whole experience, and naturalizing interactions will be key to transforming digital assistants from advertising tools into essential elements of enriched customer relationships and operational success.

\section*{Authors contributions}
Emilia Lesiak is responsible for writing the first part of the manuscript, in particular the introduction, state of the art, experiment setup, and conclusions. Grzegorz Wolny focused on the sections concerning experiment results from data analysis perspective, their discussion and also contributed to the conclusions. Bartosz Przybył provided technical details regarding the experiment setup. Finally, Michał Szczerbak helped in shaping the whole manuscript, the objectives of the experiment and conclusions.

\section*{Acknowledgements}
The experiment described in this paper was financed by the Orange Innovation Research and B2C Customer Journey departments in Orange and was technically prepared by the AI Skills Center department in Orange Innovation Poland under Paweł Tuszyński's lead. The authors wish to thank the following designers, developers, translators, and researchers who also contributed operationally to the experiment (in alphabetical order): Artur Bajll, Damian Boniecki, Mikołaj Doliński, Damian Fastowiec, Piotr Gołąbek, Maciej Jonczyk, Robert Kołodyński, Adam Konarski, Łukasz Krajewski, Anna Kraśkiewicz, Izabella Krzemińska, Valeriia Majcher, Tomasz Michalik, Filip Olechowski, Przemysław Pietrak, Robert Warzocha, Wojciech Zieliński, and Katarzyna Żubrowska.

The writing of this manuscript was supported by the use of Large Language Models for translation from Polish and for editorial-stylistic corrections only.

\bibliographystyle{unsrtnat}
\bibliography{references}

\end{document}